\documentclass{egpubl}
\usepackage{eurova2024}

\WsPaper           % uncomment for final version of EG Workshop contribution
\usepackage[T1]{fontenc}
\usepackage{dfadobe}  

\usepackage{cite}  % comment out for biblatex with backend=biber
\BibtexOrBiblatex
\electronicVersion
\PrintedOrElectronic
\ifpdf \usepackage[pdftex]{graphicx} \pdfcompresslevel=9
\else \usepackage[dvips]{graphicx} \fi

\usepackage{egweblnk} 
\usepackage{cite}
\usepackage{balance}
\usepackage{subcaption}
\usepackage{csquotes}

\usepackage{xcolor}
\newcommand{\new}[1]{#1}
\newcommand{\moved}[1]{#1}

\title[Class-Centric Visual Interactive Labeling]%
      {cVIL: Class-Centric Visual Interactive Labeling}

\author[Matthias Matt, Matthias Zeppelzauer, Manuela Waldner]
{\parbox{\textwidth}{\centering Matthias Matt$^{1}$\orcid{0009-0001-6195-3169}, Matthias Zeppelzauer$^{2}$\orcid{0000-0003-0413-4746}, Manuela Waldner$^{1}$\orcid{0000-0003-1387-5132}
        }
        \\
{\parbox{\textwidth}{\centering $^1$TU Wien, Institute of Visual Computing \& Human-Centered Technology, Austria\\
         $^2$St. Pölten University of Applied Sciences, Austria
       }
}
}
\begin{document}

\teaser{
  \includegraphics[width=\linewidth]{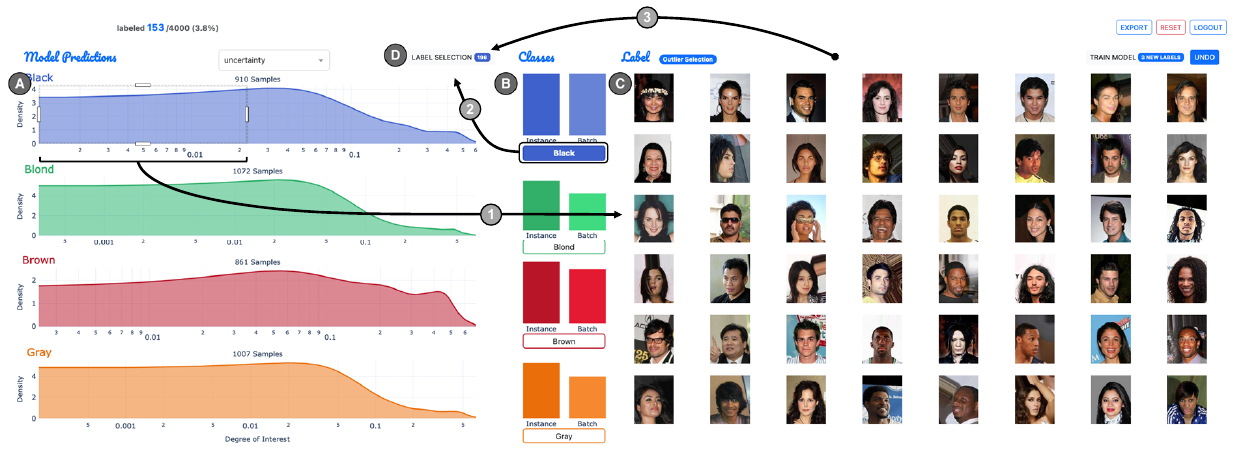}
  \centering
   \caption{Batch labeling workflow using cVIL for labeling 4,000 images according to hair color: (1) selecting a range of images with low uncertainty for black hair from the class density plot (A) yields a preview of images with the highest uncertainty in the selection (C). (2) The anticipated target class (``black'') is selected from the class distribution view (B), which shows the proportion of instances already labeled individually (\emph{instance} labels) or through a range selection (\emph{batch} labels). %Modifying the selected area in the class density plot immediately updates the sampled instances in the labeling view enabling the user to refine the selection (e.g. if the preview also shows other hair colors). 
    (3) The user confirms the target class (if all persons have black hair) by labeling all images in (C) in a batch manner (D) \new{--} or updates the selection (\new{if not all persons have black hair;} A).  
   %If a user believes a selection contains only correct samples, they can label all instances in it at once using the \textquote{Label selection} button (D). The images, whether labeled individually or through labeling a selection are assigned the selected class in the class distribution view (B). In this figure, this is the \textquote{Black} class as indicated by the highlighted button labeled \textquote{Black}. The distribution of collected labels is shown for each class through bar charts above the respective button to keep the number of labels between classes balanced. The proportion of samples labeled individually (\emph{instance} labels) or through a multi-selection (\emph{batch} labels) are shown in separate bars since these receive a different weight while training the model. Currently, for example, the user has fewer instance and batch labels for the \textquote{Blond} class.
 \label{fig:prototype}
}
}

\maketitle

%-------------------------------------------------------------------------
\begin{abstract}
   We present cVIL, a class-centric approach to visual interactive labeling, which facilitates human annotation of large and complex image data sets. cVIL uses different property measures to support instance labeling for labeling difficult instances and batch labeling to quickly label easy instances. \new{Simulated experiments reveal that cVIL with batch labeling can outperform traditional labeling approaches based on active learning.} In a \new{user} study, cVIL led to better accuracy and higher user preference compared to a traditional instance-based visual interactive labeling approach \new{based on 2D} scatterplots. 
  
%-------------------------------------------------------------------------
%  ACM CCS 1998
%  (see http://www.acm.org/about/class/1998)
% \begin{classification} % according to http:http://www.acm.org/about/class/1998
% \CCScat{Computer Graphics}{I.3.3}{Picture/Image Generation}{Line and curve generation}
% \end{classification}
%-------------------------------------------------------------------------
%https://dl.acm.org/ccs#


\end{abstract}  
%-------------------------------------------------------------------------
\section{Introduction}

Reducing costs for human annotation of unlabeled, unstructured data, \new{such as} image collections, has been widely studied. One standard methodology originating from the domain of machine learning is active learning (AL), where the model steers the labeling process~\cite{settles2009active}. One prominent AL strategy is uncertainty-based selection where the most ambiguous instance is selected for labeling \cite{Fu2012ASO}. 

Using AL, the user's role is, however, limited to labeling those instances that are autonomously selected by the system, which can easily get tedious and frustrating to users~\cite{amershi2014power}. 
A successful approach for integrating user initiative in the labeling process is visual interactive labeling (VIL). VIL has been found superior to AL at least for simple labeling tasks~\cite{Bernard2018ComparingVL}. VIL systems typically show parts or all data in a visualization, where the user can interactively select instances for labeling. For example, Seifert and Granitzer~\cite{Seifert2010UserBasedAL} show the model confidence for all instances and classes in a radial visualization. VIL has been extended by guidance mechanisms~\cite{Grimmeisen2022VisGILML} and semi-automatic labeling approaches based on clustering \cite{Chegini2019InteractiveLO, Beil2020ClustercleanlabelAI}. Some systems also allow interactively selecting a threshold based on confidence \cite{Benato2020SemiAutomaticDA} to label samples or other predefined metrics to grow clusters based on distance \cite{Schrder2020MorphoClusterEA} or purify clusters based on outlier scores \cite{Beil2020ClustercleanlabelAI}. To combine the strengths of VIL and AL, Bernard et al.~\cite{Bernard2018VIALAU} have introduced a process model, which allows both the model and user to select instances for labeling. 

\begin{figure}[htb]
     \centering
    \begin{subfigure}[t]{\linewidth}
         \centering
         \includegraphics[width=\linewidth]{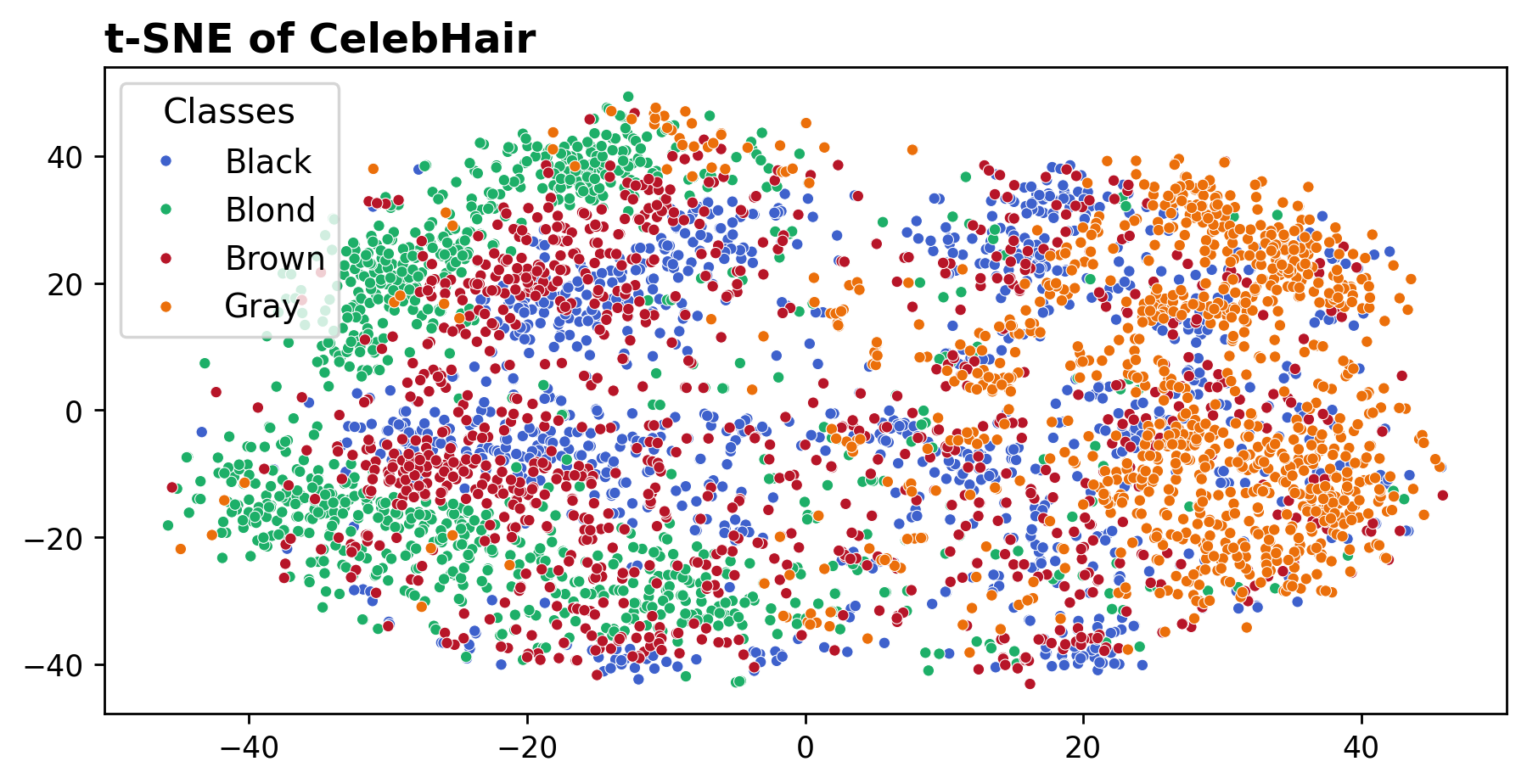}
     \end{subfigure}
    \\
    \begin{subfigure}[t]{\linewidth}
         \centering
         \includegraphics[width=\linewidth]{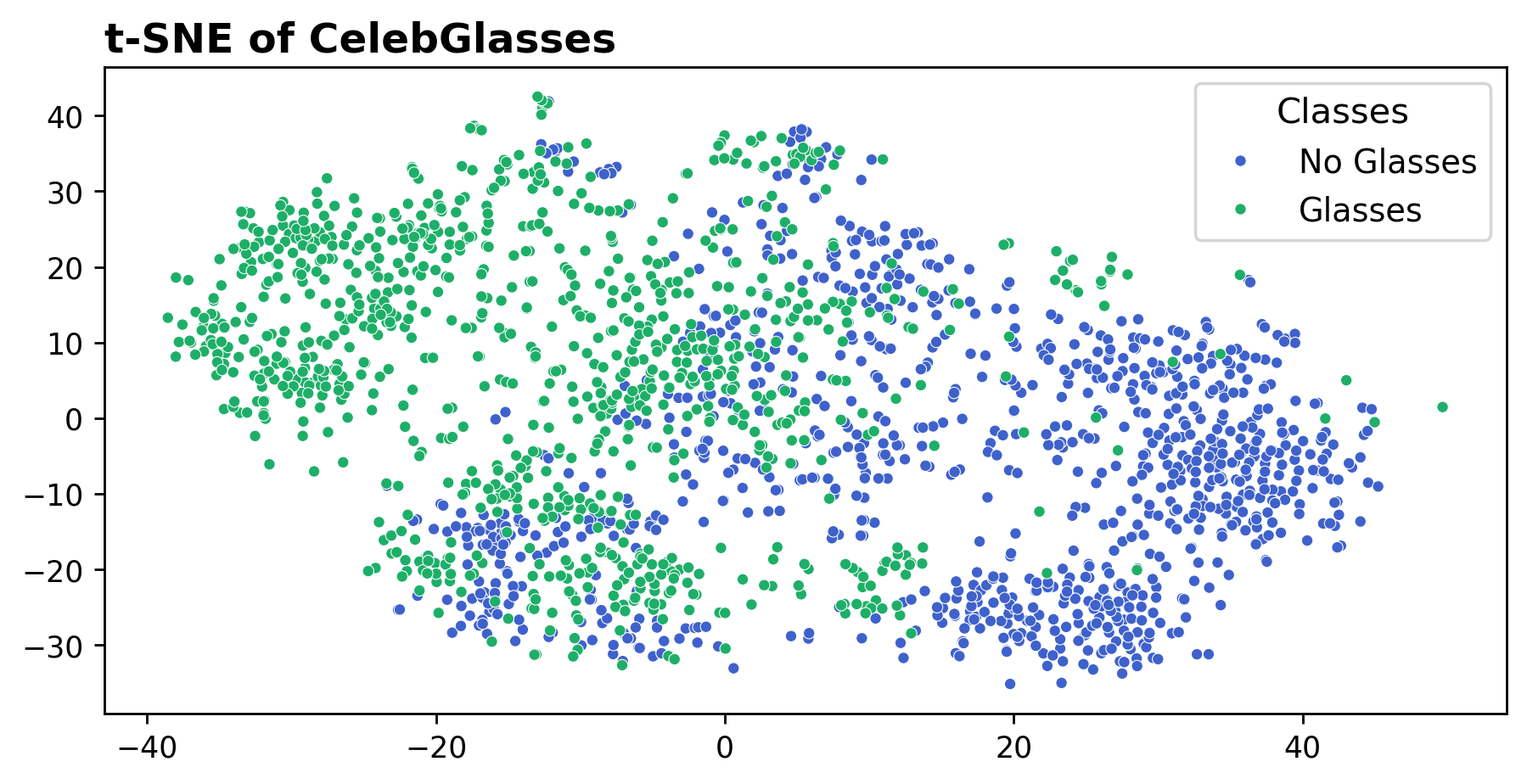}
     \end{subfigure}
     
    \caption{t-SNE projection of \new{high dimensional} embeddings found by representation learning, used for the iVIL \new{approach} in our user study.}
    \label{fig:tsne}
\end{figure}

Typically, VIL approaches, as described above, are \emph{instance-centric}, i.e., they show the class assignment of all data instances to be labeled in some spatial representation like a 2D scatterplot. We refer to such systems as iVIL. \new{When working with large amounts of data, it is impractical to interact with the data directly~\cite{Chen2011FromDA}. Consequently, }the visualization in a scatterplot can limit the system's scalability with respect to the amount of data to be visualized and labeled -- especially if the data distribution in the projection space is complex and classes strongly overlap in the projection~\cite{nonato2018multidimensional}. Furthermore, the preceding dimensionality reduction may introduce (non-linear) distortions.
Figure~\ref{fig:tsne} illustrates \new{the} limitation \new{of scatterplots}: It shows t-SNE~\cite{Maaten2008VisualizingDU} projections of two subsets of the CelebA~\cite{celeba} dataset (which we call CelebHair and CelebGlasses), where the classes refer to a person's hair color \new{(upper plot)} and whether or not they wear glasses \new{(lower plot)}. Each dataset comprises 1,000 images per class. The images are represented by DINO embeddings~\cite{Caron2021DINO}. Color encodes the ground truth class for illustration. It is easy to see that labeling this data (imagine all points would be gray because no labels exist yet) becomes a complex and tedious task. Selecting individual instances is time-consuming, and selecting multiple instances of one class at once is difficult, due to the large class overlaps.

% To address the above challenges, we propose a \emph{class-centric} VIL approach, named cVIL. The central idea of cVIL is to visualize class-wise distributions of \emph{property measures}~\cite{Bernard2021ATO} instead of individual data instances. cVIL thereby combines the advantages of AL and VIL: it computes well-understandable properties for all instances to steer the labeling process (like in AL) but ultimately leaves the initiative for selecting items to be labeled to the user (like in VIL). By showing distributions of property measures over target classes, this approach conceptually scales independently of the number of data instances and could also show more classes than users can distinguish through colors and symbols in a scatterplot. The employed density plots for representing the class-specific distributions further enable the efficient selection of batches of instances, which can significantly speed up the labeling process.

\new{
To address the above challenges, we propose a \emph{class-centric} VIL approach, named cVIL. The main contributions are:
\begin{itemize}
    \setlength{\itemindent}{.5cm}
    \item A new visual interactive labeling approach, which combines principles of AL (the computation of well-understandable properties) and VIL (leaving the initiative for selecting items to be labeled to the user; Section~\ref{sec:measures}).
    \item A labeling interface that scales independently of the number of data instances (Section~\ref{sec:encoding}) and allows for efficient instance and batch labeling (Section~\ref{sec:interaction_design}).
    \item Results from simulations that demonstrate better efficiency compared to AL (Section~\ref{sec:simulations}) and from a user study that show higher labeling accuracy and user preference compared to iVIL (Section~\ref{sec:userStudy}).
\end{itemize}
}

%-------------------------------------------------------------------------
\section{Class-Centric Visual Interactive Labeling}

% The basic principle of cVIL is to provide a visual overview of class-centric measures from which users can select instances for labeling. By showing distributions of property measures across instances at a class level rather than features per instance (like iVIL), cVIL disentangles the visual representation from the number of instances. The main objective of cVIL is to provide an interface that helps users interactively train a classifier guided by instance selection through property measures. By utilizing density plots, users can easily verify model predictions and can also label large amounts of data through batch labeling. Thus, model training, as well as the creation of labeled datasets, can be combined and parallelized.

\new{cVIL provides a visual overview of class-centric property measure values~\cite{Bernard2021ATO} from which users can select instances or batches of instances for labeling. By showing distributions of property measures across instances at a class level rather than features per instance, cVIL disentangles the visual representation from the number of instances and helps users to interactively train a classifier guided by property measures. This can speed up the labeling process considerably. }

\subsection{Measures} \label{sec:measures}

 By observing participants during an instance-centric visual labeling process, Bernard et al.~\cite{Bernard2018ComparingVL} found that participants use specific strategies while labeling. These strategies were later formalized into the concept of \emph{property measures}~\cite{Bernard2021ATO}.
 For cVIL, we experimented with three representative and complimentary property measures, which either use the model output, data distribution, or both to calculate the final result:

\textbf{Min-Margin}: This property measure is based on class probability, defining Min-Margin as the difference between the largest and second-largest class probability value. To get a measure of uncertainty, the resulting value is subtracted from 1. For example, given a class assignment $(0.1, 0.1, 0.5, 0.3)$ for a single instance, the uncertainty of this instance based on Min-Margin is given by $1-(0.5-0.3)=0.8$.

\textbf{Eccentricity}: We define sample eccentricity as the Euclidean distance to the overall median of the data. Before calculating the distance, each dimension is first scaled by the inverse of its respective variance. The variance is computed over the entire input data. This measure is completely data-centric (i.e. model-independent) in contrast to the others and enables the identification of outliers in the data distribution.

\textbf{Disagreement}: The average Jensen-Shannon distance between a data point's class probability distribution and that of all its neighbors defines its disagreement in relation to its neighborhood. The neighborhood of a sample is defined by the $k$-nearest neighbors based on Euclidean distance. While the neighbors of a sample are selected from the data space, the resulting property measure value is calculated based on the class probability distribution of the neighbors, making it a hybrid approach. A value of $k=20$ was chosen for all evaluations. This measure highlights instances in particularly inhomogeneous regions wrt.~class assignments in the feature space.

\subsection{Visual Encoding} \label{sec:encoding}

The property measures output one scalar value per unlabeled data instance, whereby each data instance is assigned to the predicted class based on the probability output of the model for this instance. The visual encoding, therefore, only needs to convey a univariate distribution per class. To ensure scalability with the number of data instances, we chose an area plot based on kernel density estimation (KDE) of the values, as shown in Figure~\ref{fig:prototype} (A). The x-axis represents the values of the respective property measure (e.g., Min-Margin), and the y-axis is the estimated density of instances corresponding to this value. The x-axes of the plots are not shared because only the relative position within each class partition is relevant for labeling.

\subsection{Interaction Design} \label{sec:interaction_design}

The KDE class distribution plots allow users to preview images by hovering or selecting images within a certain property measure range (see Figure~\ref{fig:prototype}). In the preview panel, images with the largest property measure values in the selection will then be displayed, which tend to represent the most uncertain images within the selection -- and, therefore, the most important ones for the learning and labeling process.
The number of image previews is limited to the size of the panel. Hovering over the KDE class distribution plot updates the preview images so that images with equal or lower property measure values are shown. This way, the user can quickly validate model predictions, \new{making cVIL also suitable for \emph{Quality Assurance for Machine Learning} (QA4ML)~\cite{Zhang2021SimulationBasedOO} tasks}.

After making a selection in the KDE plot and selecting a target class, users have two options for how they can assign the label of the selected target class to the images: using \emph{instance labeling}, users click on individual images from the preview to label them with the selected class. This is the most useful interaction after a cold start and when selecting images after making an outlier selection (i.e., selecting images with a high uncertainty score for a particular class) to identify and label particularly difficult instances. 

The second option is \emph{batch labeling}, which can be used to quickly confirm class predictions for instances that are rather certain. Low property measure values are a good indicator for such instances. If users think that all instances within the selected range do belong to the selected target class, they just press a button to label all instances from the current selection in the KDE plot at once, as illustrated in Figure~\ref{fig:prototype}. Note that this may lead to the labeling of more images than are currently visible in the preview panel and, therefore, may produce wrong labels for out-of-class examples. To prevent this, users can validate the selection range with the hover preview to make sure the images are correct. 
In our user study (Section~\ref{sec:userStudy}), we did not observe that batch labeling led to a critical number of incorrect labels. 
After model retraining, \new{which took an average of three seconds in our user study,} all KDE plots are updated with the new property measure values. \new{Due to the relatively short re-training time, users can easily update the model while labeling.}

%-------------------------------------------------------------------------
\section{Quantitative Experiments: cVIL vs.~AL} \label{sec:simulations}

\new{Firstly, we compared the property measures in terms of their runtime.}
\moved{
When benchmarking our implementation of the three property measures with an M1 MacBook Pro, we found that -- for an embedding size of 1,024 dimensions and up to more than 16,000 samples -- the property measures for all instances can be calculated in less than a second. In particular, the results for Disagreement took close to one second to calculate, while Eccentricity took less than 100ms and Min-Margin only around 1ms.
}

To assess the performance of the selected property measures and the potential of cVIL, we compared the theoretical performance of cVIL (accuracy per simulated labeling step) to AL. 
For evaluation we use MNIST \cite{lecun2010mnist}, STL10 \cite{coates2011stl10}, as well as CelebHair and CelebGlasses. For MNIST, the raw pixels are used as training data. For the more complex datasets, we use DINO~\cite{Caron2021DINO} for representation learning for the simulations as well as the user study. 
The backbone model of DINO is a vision transformer, which was pre-trained on ImageNet \cite{Deng2009ImageNetAL}.

\new{We simulate instance labeling by selecting samples with the highest property measure value in each class partition and assigning them their correct ground truth label. An equal number of samples are selected from each class to get a total of ten samples per iteration.} For example, for STL10, one sample with the highest property measure value per class is selected. We simulate optimal batch labeling by selecting instances with the lowest property measure value up to the first incorrectly predicted instance for labeling. Therefore, the number of samples in a batch selection can vary, depending on the accuracy of the model and property measures.

% Figure \ref{fig:schema} shows the steps for one labeling iteration. We simulate instance labeling by selecting a total of 10 samples per iteration and assigning them their correct ground truth label. The selection is balanced where an equal number of samples is selected from each class. For example, for STL10, one sample with the highest property measure value per class is selected. We simulate optimal batch labeling by selecting instances with the lowest property measure value up to the first incorrectly predicted instance for labeling. Therefore, the number of samples in a batch selection can vary, depending on the accuracy of the model and property measures.

% \begin{figure}[htb]
%     \centering
%     \includegraphics[width=\linewidth]{assets/labeling_simulation_schema.png}
%     \caption{Simulation step of CelebHair (black and gray hair): Instance labeling selects a total of ten instances (red boxes) while batch labeling labels a variable number of images (orange boxes). }
%     \label{fig:schema}
% \end{figure}

The classifier is a multi-layer perceptron with two hidden layers consisting of 50 and 20 neurons. Batch labels are given a lower sample weight during training, which is realized by assigning them lower costs in the loss function before back-propagation. The number of batch labels that are used for each class during training is determined by the number of samples $C_{min}$ in the class with the fewest instance labels. In the evaluations, $10 \cdot C_{min}$ instances were then randomly sampled from the batch labels of each class. The weights of samples labeled through batch labeling were set to 0.1.

\begin{figure}[t]
     \centering
    \begin{subfigure}[t]{\linewidth}
         \centering
         \includegraphics[width=\linewidth]{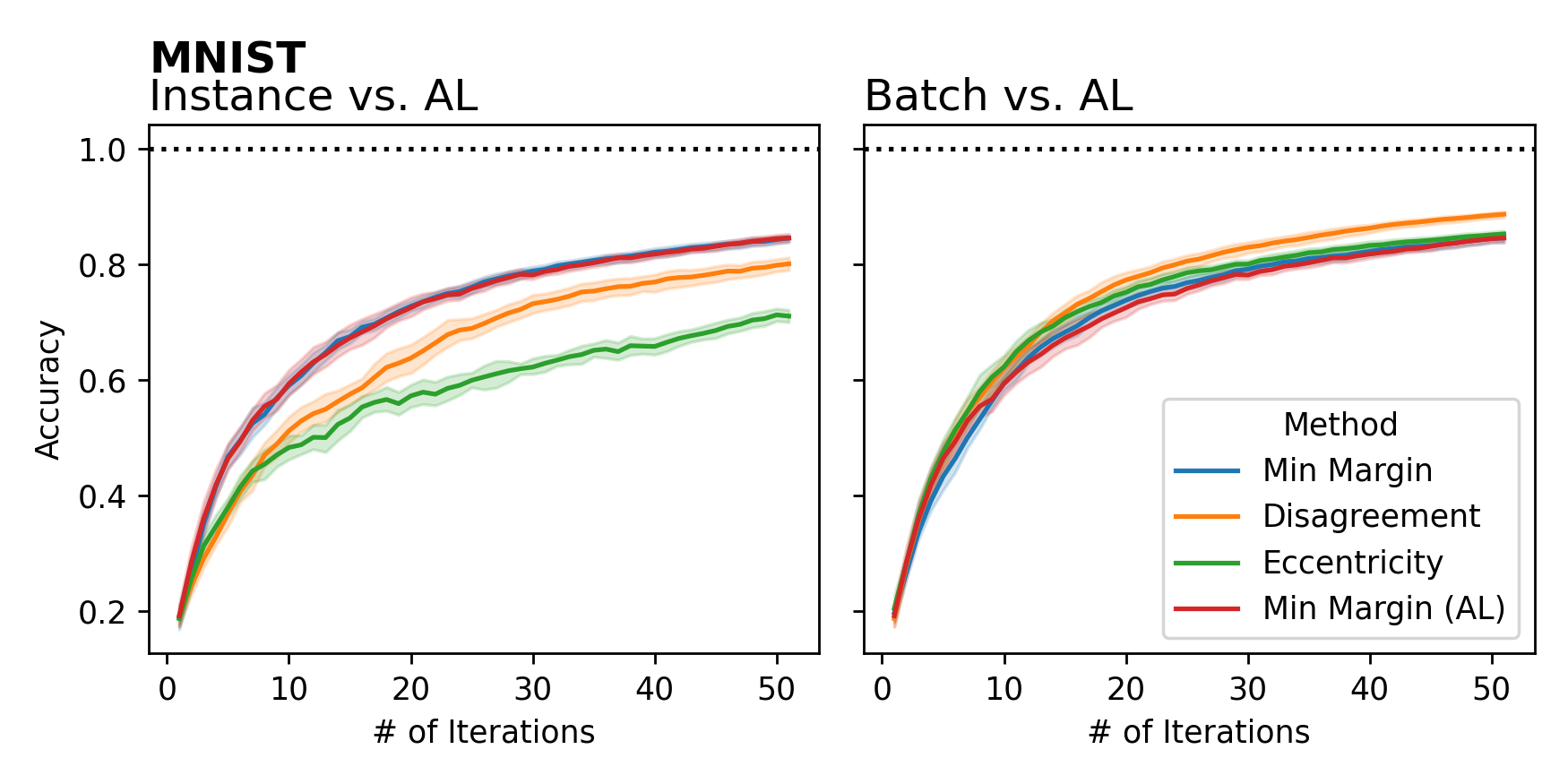}
     \end{subfigure}
    \\
    \begin{subfigure}[t]{\linewidth}
         \centering
         \includegraphics[width=\linewidth]{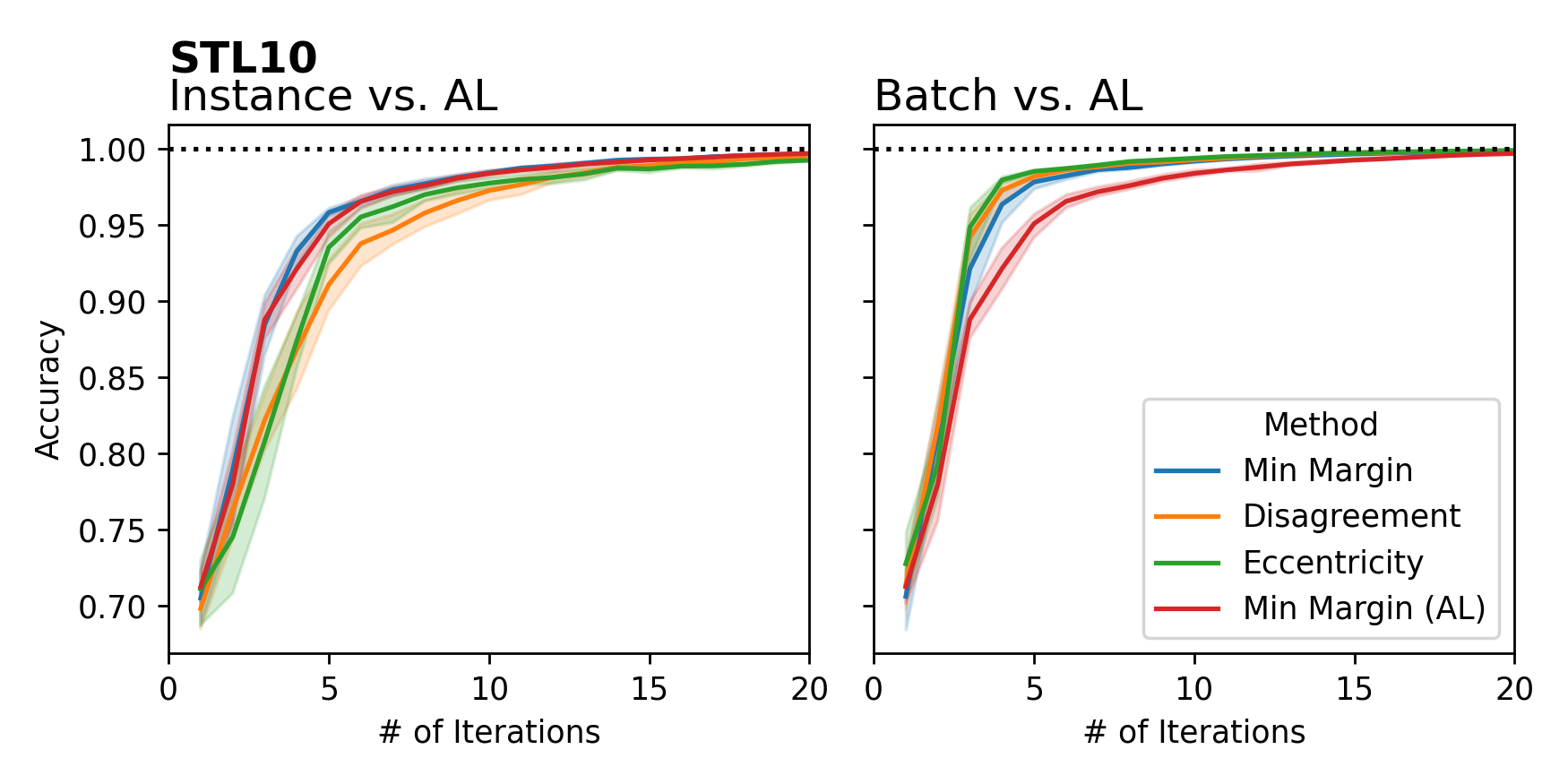}
     \end{subfigure}
     
    \caption{cVIL with Min-Margin is en par with AL for instance labeling (left). For batch labeling, cVIL outperforms AL independently of the property measure (right). %Note that we only show the simulation up to 20 iterations for STL10. 
    }
    \label{fig:al_vs_lava}
\end{figure}

Based on the evaluation results with MNIST and STL10 datasets, shown in Figure \ref{fig:al_vs_lava}, we observed that Disagreement and Eccentricity are less effective for instance labeling. Min-Margin leads to faster model improvement and is en par with AL simulations that also use Min-Margin. This indicates that class partitioning with instance labeling is comparable in performance to AL. For STL10, Eccentricity shows slightly better results than Disagreement, while the opposite is true for MNIST. For CelebHair and CelebGlasses, Eccentricity has a considerably lower accuracy, while all other methods -- including AL -- perform similarly. To summarize, instance labeling using cVIL performs best with Min-Margin and yields comparable performance to Min-Margin-based AL.

For batch labeling, we observe that Disagreement and Eccentricity slightly outperform Min-Margin for MNIST and both strategies outperform AL. For CelebHair and CelebGlasses, all methods perform similarly well. For STL10, which is the most complex dataset, cVIL clearly outperforms AL with all three property measures, with peak performance for Eccentricity. 
To sum up, batch labeling can easily compete with AL and outperforms it, in particular for complex data, even when using the same measure (Min-Margin). Surprisingly, a purely data-driven strategy (Eccentricity) outperforms all others. Our experiments show that cVIL with batch labeling has clear potential to outperform traditional AL.

\section{User Study: cVIL vs.~iVIL}
\label{sec:userStudy}

In the user study, we compare our class-centric VIL interface (cVIL) to a simple instance-centric VIL interface (iVIL). For the iVIL baseline interface, the KDE plots from Figure \ref{fig:prototype} (A) are replaced with a t-SNE~\cite{Maaten2008VisualizingDU} projection of the DINO feature vectors~\cite{Caron2021DINO}, as shown in Figure \ref{fig:tsne}. The colors shown in the interface represent the \emph{predicted} labels associated with the data instances.
Users can either label individually selected instances or perform batch labeling using a lasso selection in the scatter plot. 
Our hypothesis is that cVIL is more effective at labeling than iVIL while being less demanding and more preferable to use.

\subsection{Study Design}

\textbf{Participants}: The user study was conducted with 16 participants (11 male, 4 female, 1 non-binary). All of them had a background in computer science, with 4 being post-graduate, 6 graduate, and 6 undergraduate level. Moreover, 11 of the participants had prior experience with machine learning. All participants were between 22 and 40 years old with a median age of 26.

\textbf{Task}: During the study, the participants were given labeling tasks. The system was not initialized, which meant that participants had to first label samples based on a random selection to train the model at which point the full interface became available. The task was considered complete only when participants believed that all the data points had their correct label -- either assigned manually by the user or predicted by the model. The participants had to decide when this stage was reached and were instructed to export the labels at that point. The entire study, including setup and introduction, was estimated to take approximately one hour to complete. 

\textbf{Data}: We asked users to label the CelebGlasses and CelebHair data sets (see Figure~\ref{fig:tsne}). To reduce the complexity of the CelebHair dataset, a subset with only \new{black} and gray hair was used. To reduce the amount of time required for labeling, the number of samples was limited to 1,000 per class. For the user study, only Min-Margin was available to the participants due to its simplicity and good overall performance. This further reduced the task complexity by only presenting users with a single measure to focus on. 

\textbf{Independent and dependent variables}: The study used a within-subjects design, with the interface (cVIL vs.~iVIL) as the independent variable. We also varied the dataset assigned to the two interfaces, as well as the order of appearance of the interfaces. The overall accuracy of both manual and predicted labels, labeling time, cognitive demand, and preference were the dependent variables being observed. 

\textbf{Procedure}: Participants were provided with a tutorial sheet that explained the system's components and how to interact with them prior to attempting the task. They solved the tasks on an external monitor and had access to a mouse and keyboard. The study was conducted on an M1 MacBook Pro. After solving each task, the participants completed a NASA TLX questionnaire to evaluate the perceived cognitive demand of the tasks. Upon completing the study, participants were asked to express their likes and dislikes about each interface and indicate their preference.

\subsection{Results}

Of the 16 participants, 15 successfully finished the study. One user accidentally selected an incorrect target class for a batch labeling action using cVIL, which led to a dramatic reduction in accuracy. We excluded this participant from further analysis (see Section~\ref{sec:discussion} for our learnings from this case). 

\begin{figure}
\begin{minipage}[t]{0.45\linewidth}
        \centering
        \includegraphics[width=\linewidth]{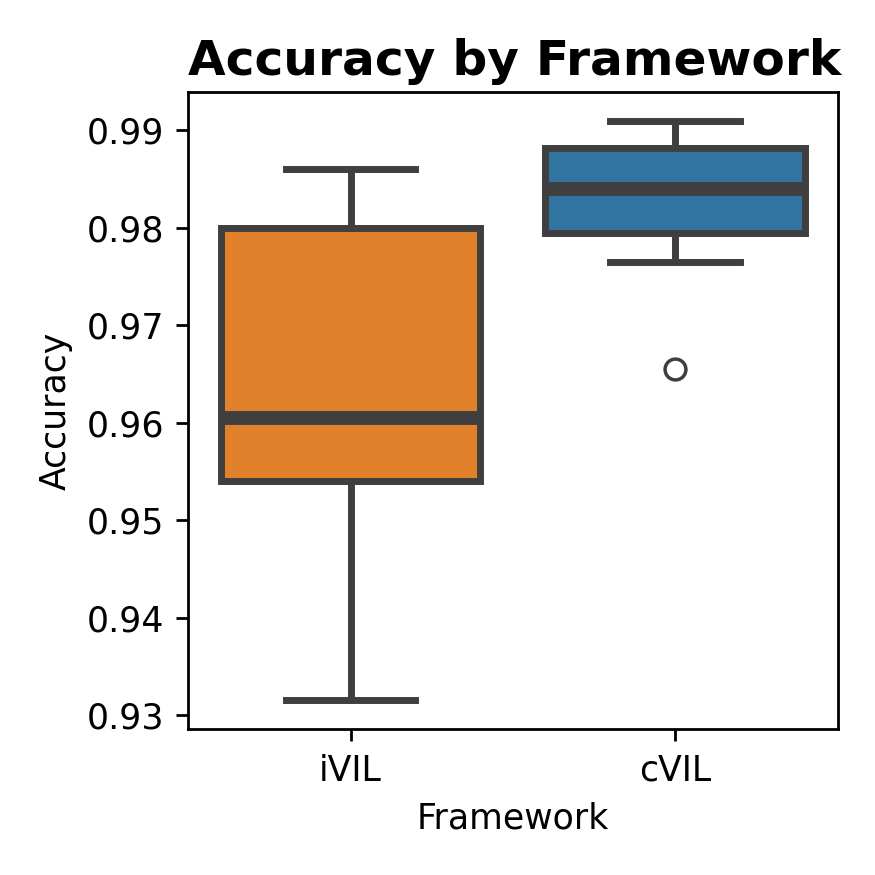}
        \caption{ Final accuracy.  }
        \label{fig:individual_results}
\end{minipage}
\hfill
\begin{minipage}[t]{0.45\linewidth}
        \centering
        \includegraphics[width=\linewidth]{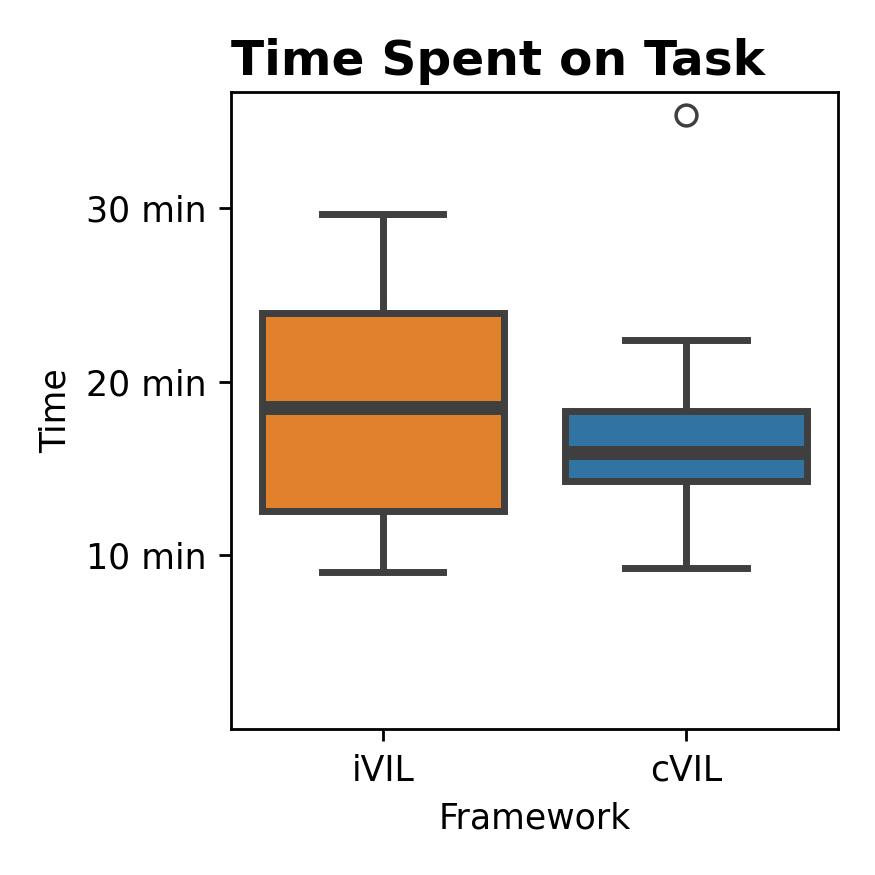}
        \caption{ Completion time. }
        \label{fig:labeling_time}
\end{minipage}%
\end{figure}

All remaining participants performed better using cVIL compared to iVIL. The median accuracy of the exported labels compared to the ground truth was 96.05\% for iVIL, whereas it was 98.4\% for cVIL, as can be seen in Figure \ref{fig:individual_results}, which is a statistically significant difference ($t(14)=5.784$, $p<.001$).

\begin{figure}
\begin{minipage}[t]{0.45\linewidth}
        \includegraphics[width=\linewidth]{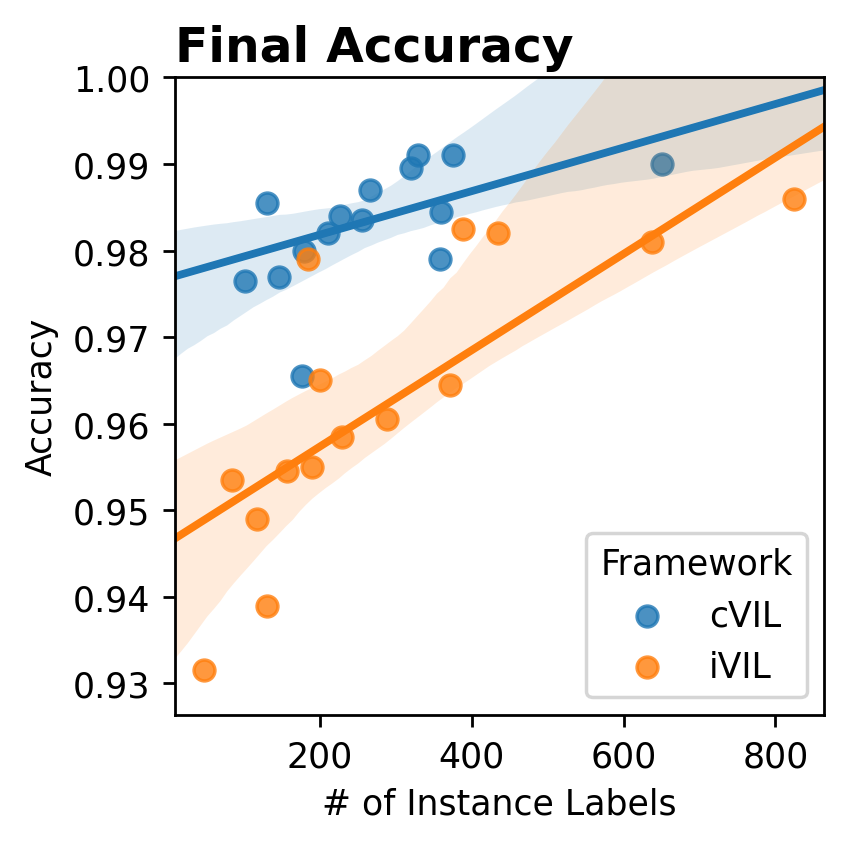}
        \caption{ Final accuracy by number of instance labels.  }
        \label{fig:lm}
\end{minipage}
\hfill
\begin{minipage}[t]{0.45\linewidth}
        \includegraphics[width=\linewidth]{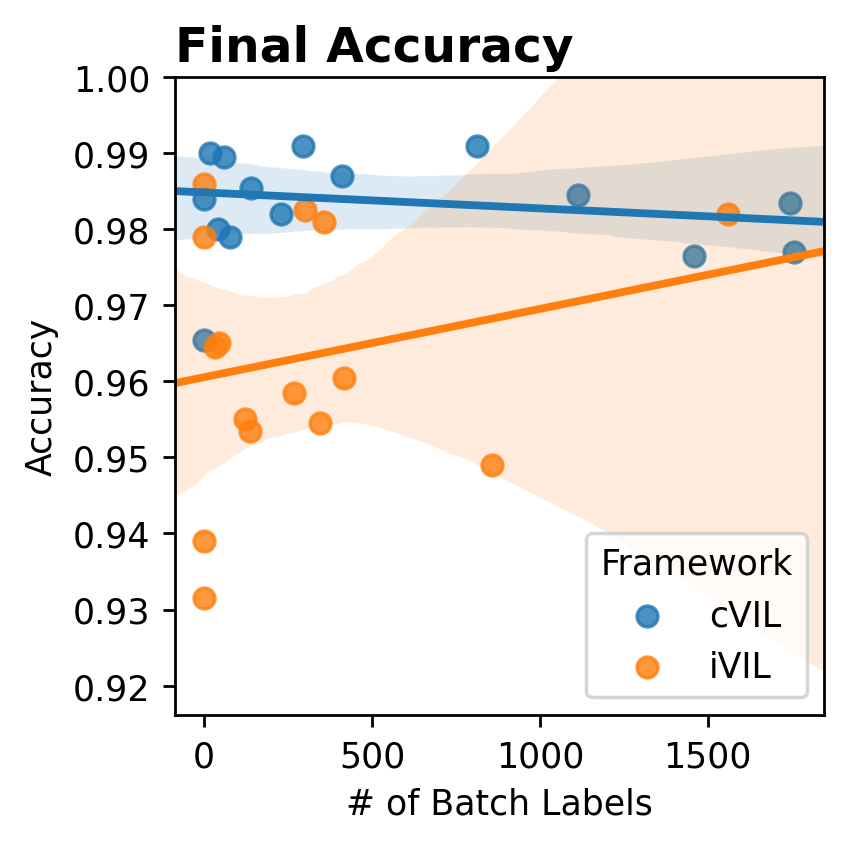}
        \caption{ Final accuracy by number of batch labels.  }
        \label{fig:lm_batch}
\end{minipage}%
\end{figure}

When comparing the final accuracy (measured against the ground truth) in dependence on the number of instance labels, we observe that cVIL achieves considerably better accuracy with fewer labels, as can be seen in Figure \ref{fig:lm}. The solid lines represent a robust linear regression estimation of the results for each framework. cVIL achieves around the same accuracy with 100 labels as iVIL with 600.
Notably, these results were achieved solely through instance labeling, as can be seen in Figure \ref{fig:lm_batch}. Interestingly, the accuracy is not significantly affected by the number of generated batch labels in both conditions  (except for one outlier in iVIL).
Participants batch-labeled an average of 600 samples in cVIL and 315 in iVIL, but this difference in the number of labels is not statistically significant ($t(14)=1.570$, $p=0.14$). 

We also measured the time it took to finish the labeling task by looking at the difference between the first and the last labeling action or model retraining. Participants generally needed less time to finish the labeling tasks in cVIL as can be seen in Figure \ref{fig:labeling_time}. The median labeling time for iVIL was around 18:30 minutes compared to 16 minutes for cVIL, however, this difference is not statistically significant ($t(14)=-1.947$, $p=0.07$).

To analyze task load, we aggregated the scores from the NASA TLX questionnaire as described by Rubio et al.~\cite{Rubio2004EvaluationOS}, where we weighted Frustration \new{highest}, followed by Mental Demand, Effort, Performance, and finally Temporal Demand. We could detect no difference between cVIL and the baseline (Wilcoxon Signed-Rank test: $z=-0.879; p=0.39$).

In the final questionnaire, however, 13 out of 16 participants stated that they preferred cVIL over the iVIL baseline implementation. The main reason for this was that participants found the system easier to use as it helped them to better spot false positives and outliers, evaluate current model performance, and re-label incorrect predictions. Having information about uncertainty made it easier to label and gave a better indication of the model's accuracy. Participants reported that the class partitioning made it easier to focus on one particular class, and participants only needed to identify false positives when deciding if a class label was correct, rather than having to consider the possibility of having true and false negatives as in an instance-based approach. However, participants disliked that there was too little change in the visual representations after the model updates. Four participants reported that the iVIL baseline was also easier to understand and navigate as well as more engaging and fun to use. However, six other participants felt it was more tedious and ambiguous since images were harder to find.

%-------------------------------------------------------------------------

\section{Discussion and Conclusions}
\label{sec:discussion}
In our study, users could label 2,000 photos in a reasonable time and with high accuracy with both interfaces. All users could achieve higher performance using cVIL, and most participants preferred it over the instance-based baseline. This confirms that class-centric VIL is promising when data is getting more complex -- even though the data used for our user study was still relatively small. We conclude that scatterplots provide a less reliable way to judge model performance when class distributions strongly overlap compared to our class-centric property measure distribution plots. We found that cVIL can address these limitations and provide effective feedback with large and complex datasets while still being fully interactive.

Our study showed several directions for improvements of cVIL. In our study, we had to exclude one user because of an incorrect target class selection for batch labeling, which had drastic consequences on model accuracy. This could have happened with both interfaces and could be easily prevented for cVIL by warning the user if the selected target class does not correspond to the class density plot from which the selection was made. Generally, batch labeling using Min-Margin as the property measure (and also lasso selections in the baseline) did not notably contribute to model accuracy. In our simulations, data-based property measures produced more promising results for batch labeling. In the future, the usage of different property measures for instance and batch labeling should be investigated and empirically evaluated. Both our simulations and the user study show that cVIL is a promising approach in cases where data complexity becomes high and sample sizes are large, as is typically the case in real-world scenarios. Therefore, focusing on one class at a time and enabling the user to switch between instance and batch labeling depending on the state of the learning process shows to be of central importance.

A number of open topics remain for future investigation, such as the choice of the most suitable property measure, investigating different batch labeling strategies (including methodology from semi-supervised learning), and systematically evaluating the impact of data set size, features, and class properties on instance- and class-centric VIL performance.

\section*{Acknowledgements}

\new{We thank Professor Takeo Igarashi for valuable discussions and all participants of the user study for their time.}

\new{This research was funded in whole or in part by the Austrian Science Fund (FWF) \href{https://doi.org/10.55776/P36453}{10.55776/P36453}. For open access purposes, the author has applied a CC BY public copyright license to any author accepted manuscript version arising from this submission.}

%-------------------------------------------------------------------------
\balance
\bibliographystyle{eg-alpha-doi}

\bibliography{egbibsample}

%-------------------------------------------------------------------------
\newpage

\end{document}